\documentclass[12pt,preprint]{aastex}
\usepackage{lineno}
 \linenumbers*[1]

\newcommand{\be}{\begin{equation}}
\newcommand{\ee}{\end{equation}}
\newcommand{\bea}{\begin{eqnarray}}
\newcommand{\eea}{\end{eqnarray}}

\shorttitle{}
\shortauthors{Col\'on \& Gaidos}

\begin{document}
\title{Narrow-$K$-Band Observations of the GJ 1214 System}

\author{Knicole D.\ Col\'on\altaffilmark{1}, Eric Gaidos\altaffilmark{1}}

\altaffiltext{1}{Department of Geology and Geophysics, University of Hawai'i at Manoa, Honolulu, HI 96822; colonk@hawaii.edu}

\begin{abstract}
GJ 1214 is a nearby M dwarf star that hosts a transiting super-Earth-size planet, making this system an excellent target for atmospheric studies.  Most studies find that the transmission spectrum of GJ 1214b is flat, which favors either a high mean molecular weight or cloudy/hazy hydrogen (H) rich atmosphere model.  Photometry at short wavelengths ($<$ 0.7 $\mu$m) and in the $K$-band can discriminate the most between these different atmosphere models for GJ 1214b, but current observations do not have sufficiently high precision.  We present photometry of seven transits of GJ 1214b through a narrow $K$-band (2.141 $\mu$m) filter with the Wide Field Camera on the 3.8 m United Kingdom Infrared Telescope.   Our photometric precision is typically 1.7$\times$10$^{-3}$ (for a single transit), comparable with other ground-based observations of GJ 1214b.  We measure a planet-star radius ratio of 0.1158$\pm$0.0013, which, along with other studies, also supports a flat transmission spectrum for GJ 1214b.  Since this does not exclude a scenario where GJ 1214b has a H-rich envelope with heavy elements that are sequestered below a cloud/haze layer, we compare $K$-band observations with models of H$_2$ collision-induced absorption in an atmosphere for a range of temperatures.  While we find no evidence for deviation from a flat spectrum (slope $s$ = 0.0016$\pm$0.0038), an H$_2$ dominated upper atmosphere ($<$ 60 mbar) cannot be excluded.  More precise observations at $<$ 0.7 $\mu$m and in the $K$-band as well as a uniform analysis of all published data would be useful for establishing more robust limits on atmosphere models for GJ 1214b.
\end{abstract}
\keywords{planetary systems --- planets and satellites: atmospheres --- techniques: photometric}

\section{Introduction}
\label{intro}
To date, approximately 300 planets have been confirmed to transit their host star(s), and the $Kepler$\footnote{http://kepler.nasa.gov/} mission has discovered over 3500 additional transiting planet candidates.  About 65 transiting planets (which have both a measured mass and radius) are considered to be in either the ``super-Earth'' (1 $\lesssim$ $R$ $\lesssim$ 2 $R_{\oplus}$) or ``mini-Neptune'' (2 $\lesssim$ $R$ $\lesssim$ 4 $R_{\oplus}$) regime, but less than a handful of these orbit nearby stars \citep{schneider2011}.  Planets like GJ 1214b, a $\sim$ 2.7 $R_{\oplus}$ transiting planet discovered around a nearby $\sim$ 0.2 $R_{\odot}$ M dwarf star by the MEarth ground-based transit survey \citep{charb2009}, are therefore of great interest for understanding the difference between Earth-like planets, ``super-Earths'', and ``mini-Neptunes.''  

Planets in the super-Earth/mini-Neptune regime occupy a location in the planetary mass-radius diagram that allows for diverse interior compositions \citep{rogers2010}.  However, it may be possible to constrain the overall bulk composition of these planets by characterizing the planet's atmosphere.  Models suggest that GJ 1214b contains a significant amount of hydrogen (H) and helium (He) based on its low mean density \citep[approximately one-third that of Earth;][]{miller2010}.  \citet{valencia2013} constrained the fraction of H and He in GJ 1214b using a model of its interior and evolution.  They conclude that there is some amount of H/He present and that the bulk amount of H/He may be up to 7\% by mass (similar to Neptune).  While this is only a small fraction, it suggests that GJ 1214b likely has \emph{some} amount of H/He in its atmosphere.  Since H has a large scale height due to its low molecular weight, even with a small amount of H in the atmosphere the upper atmosphere (i.e. a small distance above the homopause) can easily be H-dominated \citep{pierre2010}.  Thus, if GJ 1214b is differentiated, then ice and rock will be concentrated in the interior and the concentration of H/He in the surrounding atmosphere could be much higher than 7\%.
 
\begin{figure}
\begin{center}
{\includegraphics[scale=0.6, angle=90]{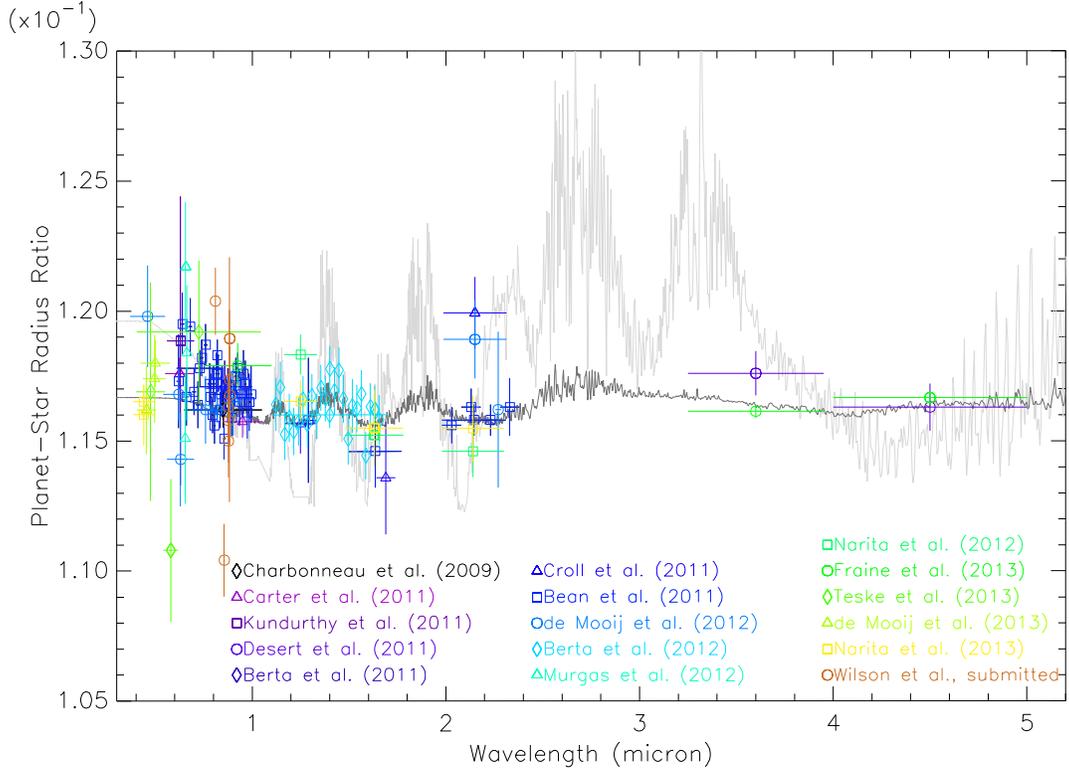}} 
\end{center}
\caption{$R_p$/$R_{\star}$ from published observations of GJ 1214b.  The horizontal error bars on each point indicate the approximate bandpass for each observation.  The data points are color-coded according to the source they were retrieved from and are shown in order of publication date.  Also shown are two atmosphere models from \cite{howe2012} that best fit a majority of the data published by 2012.  The light gray line is a model for a solar composition atmosphere with a tholin haze at 10-0.1 mbar composed of 0.1 $\mu$m particles with a particle density of 100 cm$^{-3}$.  The dark gray line is a model for an atmosphere with a composition of 1\% H$_2$O and 99\% N$_2$ and with a tholin haze at 0.1-0.001 mbar composed of 0.01 $\mu$m particles with a density of 10$^6$ cm$^{-3}$.    
\label{specall}}
\end{figure}

Many groups have studied its atmosphere by transmission spectroscopy or spectrophotometry \citep[e.g.,][]{bean2010,bean2011,berta2011,croll2011,cross2011,desert2011,berta2012,demoo2012,murgas2012,narita2012,fraine2013,demoo2013,narita2013,teske2013} while others have continued improving models of its atmosphere \citep[e.g.,][]{benn2012,heng2012,howe2012,menou2012,kempton2012,morley2013,valencia2013}.  Figure \ref{specall} presents measurements of the planet-star radius ratio (or simply, the radius ratio, $R_p$/$R_{\star}$) of GJ 1214b, along with two representative atmosphere models from \citet{howe2012}.  GJ 1214b has a largely flat, featureless spectrum, which supports an atmosphere with a high molecular weight and small scale height, and/or a strongly scattering layer (clouds or aerosols) \citep[e.g.,][]{berta2012}.  However, published observations at $<$ 0.7 $\mu$m and in the $K$-band ($\sim$ 1.9$-$2.5 $\mu$m) as yet cannot discriminate between these two scenarios.  Different observations from \citet{bean2011}, \citet{carter2011}, \citet{kund2011}, \citet{demoo2012}, \citet{murgas2012}, \citet{narita2013}, and \citet{demoo2013} disagree as to whether or not there is a rise in the spectrum at $<$ 0.7 $\mu$m due to Rayleigh scattering.  In the $K_s$-band, observations from \citet{croll2011} have been interpreted as showing a deviation from a flat spectrum, suggesting a lower mean molecular weight atmosphere, while \cite{bean2011}, \citet{demoo2012}, \citet{narita2012}, and \citet{narita2013} find a flat spectrum.  The disagreement could be a result of the use of slightly different bandpasses combined with telluric effects at the edges of the $K$ passband, so a $K$-band filter that avoids the edges of the bandpass could help resolve the disagreement between these observations and help determine whether GJ 1214b has a high mean molecular weight atmosphere or a cloudy/hazy H-rich atmosphere. 

In one of the latest of many model analyses of the data, \citet{howe2012} presented a suite of atmosphere models for GJ 1214b.  They ultimately selected five models that they deemed to best fit the published data, with three models of a solar-abundance atmosphere (two with hazes with different particle sizes and densities and one with a uniformly opaque cloud layer) and two of an atmosphere of 1\% H$_2$O and 99\% N$_2$ plus either haze or no haze.  They ruled out several models including a H-rich atmosphere with no haze, a H-rich atmosphere with a haze of smaller ($\sim$ 0.01 $\mu$m) tholin particles, as well as a H-poor atmosphere with major sources of absorption other than water.  The model that best fits the short-wavelength rise (at $<$ 0.7 $\mu$m; Figure \ref{specall}) is a solar composition atmosphere with a (somewhat arbitrary) tholin haze layer having a particle size of 0.1 $\mu$m and extending over pressures of 10-0.1 mbar.  This model is also the best fit to the $K$-band data, but only if the observations from \citet{croll2011} are correct over the observations from \cite{bean2011}, \citet{demoo2012}, \citet{narita2012}, and \citet{narita2013}, since a flat spectrum in the $K$-band is inconsistent with a rise at short wavelengths.  \citet{howe2012} conclude that if the rise at short wavelengths is valid, GJ 1214b should have a H-rich atmosphere (albeit with some cloud or haze layer) rather than being composed primarily of heavier molecules like water or nitrogen.

In this paper, we investigated the scenario of a H-rich atmosphere with a high cloud/aerosol layer using narrow-$K$-band observations of transits of GJ 1214b.  We present observations of seven transits of GJ 1214b through a narrow-band filter centered at the 2.141 $\mu$m 1-0 S(1) vibrational line of H$_2$ (hereafter, referred to simply as the H$_2$ filter).  We used our observations to (1) help resolve the disagreement between $K$-band measurements published to date and (2) test if there is any spectral structure in the $K$-band that was missed by the broad-band observations.  Specifically, we (1) compared published $K$-band data to test if the disagreement between $K$-band measurements is the result of subtle differences in the bandpasses used convolved with not-so-subtle differences in the spectrum of the planet and telluric absorption/emission that were not previously appreciated, and (2) compared $K$-band data with a model of H$_2$ collision-induced absorption in an atmosphere for a range of temperatures.  

We describe our observations and data reduction in Section \ref{obs}.  In Section \ref{analysis}, we present our light curve analysis and models.  We present our results in Section \ref{results}, and in Section \ref{discuss} we compare our results with published $K$-band observations, compare the data with models of a H-rich atmosphere, investigate variability in the stellar spectrum due to H$_2$, and discuss the effects of different systematics on our results.  In Section \ref{conc}, we summarize our conclusions and offer suggestions for future work.  

\section{Observations and Data Reduction}
\label{obs}
We acquired photometry of seven transits of GJ 1214b between August 2011 and July 2012 with the Wide Field CAMera (WFCAM) on the 3.8 m United Kingdom InfraRed Telescope (UKIRT) \citep{cas2001}, located at the Mauna Kea Observatory in Hawaii.  WFCAM is a near-infrared wide field imager consisting of 4 Rockwell Hawaii-II (HgCdTe 2048$\times$2048) 0.4 arcsec pixel arrays.  Each individual camera covers a field of 13.65$^{'}$$\times$13.65$^{'}$, and the total field of view (FOV) is 0.75 square degrees.  A narrow-band filter centered on the fundamental H$_2$ vibrational line [S(1) $\nu$ = 1 $\rightarrow$ 0] at 2.141 $\mu$m (FWHM = 0.021 $\mu$m) was used for all observations.  This filter probes wavelengths at which previous observations are in disagreement (Section \ref{intro}) and minimizes systematic variations from Earth's atmosphere.  In Figure \ref{trans}, we compare the transmission profile of the H$_2$ filter to the broad-band $K_s$ filter [the same filter used by \citet{croll2011} on the Wide-field InfraRed Camera on the Canada-France-Hawaii Telescope] and the atmosphere above Mauna Kea Observatory (at an airmass of 1 and with a water vapor column of 1.2 mm).\footnote{These data, produced using the program IRTRANS4, were obtained from the UKIRT worldwide web pages; http://www.jach.hawaii.edu/UKIRT/astronomy/utils/atmos-index.html}  Because it is so narrow, the H$_2$ filter is an optimal filter to avoid atmospheric effects (compared to the broader $K_s$ filter).  

In Table \ref{obstable}, we present details of each transit observation.  All observations were performed in service mode.  An exposure time of 60 s was used for each integration, with a typical dead time between exposures of $\sim$2-3 s.  The observation epochs (at mid-exposure) were extracted from the FITS header for each image.  The telescope was intentionally defocused to avoid saturation and to spread the stellar PSF over many pixels to minimize error from an imperfect flat-field correction of the detector response.  Due to the amount of defocus, the telescope auto-guider was not able to function properly.  As a result, a drift of 45-52 pixels in the position of the target centroid occurred during the 2011 observations.  For the 2012 observations, the telescope operator routinely adjusted the guider to keep the defocused guide star centered in the guider acquisition window.  A centroid drift of less than 9 pixels was maintained during the 2012 observations.

We reduced all images using software written in GDL.\footnote{GNU Data Language; http://gnudatalanguage.sourceforge.net/}  While all images taken with WFCAM are processed through a pipeline operated by the Cambridge Astronomical Survey Unit (CASU), we opted to process our images separately to ensure accurate calibration and photometry that was as precise as possible.  The procedures were illumination correction, dark current correction, and flat fielding prior to performing circular aperture photometry, sky subtraction, and finally a radial distortion correction.

The illumination correction rectified each image for residual systematics, which are most likely caused by either low-level non-linearity in the detectors, scattered light in the camera, and/or spatially dependent PSF corrections.\footnote{http://apm49.ast.cam.ac.uk/surveys-projects/wfcam/technical/photometry}  Illumination correction tables are measured monthly as a function of spatial location in the array.  Since illumination correction tables are only available for broad-band filters, we used the $K$-band tables as a proxy for the H$_2$ filter.  For the dark current and flat-field correction, we used darks taken the night of each observation and monthly twilight flats taken with the H$_2$ filter.  The appropriate flats from a given month were used.  Circular aperture photometry was performed with radii = 8, 10, 12, 14, 16, 18 pixels (3.2, 4.0, 4.8, 5.6, 6.4, 7.2 arcsec).  In our analysis we ultimately used the aperture that gave the best photometric performance (Section \ref{lcs}).  An annulus of 25 - 30 pixels (10 - 12 arcsec) was used for sky subtraction.

Finally, a radial distortion correction was applied to the sky-subtracted flux measured within a given aperture to account for non-negligible field distortion in WFCAM, a result of its extremely large FOV.  Specifically, photometry of sources near the edge of the FOV have a systematic error of up to 0.02 mag.  We computed the corrected flux, $F_{corr}$, from 

\be
F_{corr}=\frac{F(1+k_{3}r^2) }{1+3k_{3}r^2},
\ee

where $F$ is the sky-subtracted flux measured within a given aperture, $k_{3}$ is the coefficient of the third order polynomial term in the radial distortion equation and is approximately -60 radian$^{-2}$ in the $K$-band \citep[][and references therein]{hodgkin2009}, and $r$ is the distance of a star relative to the center of the optical system in radians.  The corrected flux for the target and each comparison star was used to generate the light curves (Section \ref{analysis}). 

\section{Analysis}
\label{analysis}

\subsection{Selection of Comparison Stars}
\label{refs}
Thanks to the large FOV covered by WFCAM, we could select among numerous comparison stars for relative photometry.  We selected comparison stars that are of similar brightness to GJ 1214 and that do not appear to be intrinsically variable.  We also used colors to select dwarf stars over giants and that are as close as possible to GJ 1214 in spectral type.

\begin{figure}
\begin{center}
{\includegraphics[scale=0.6, angle=90]{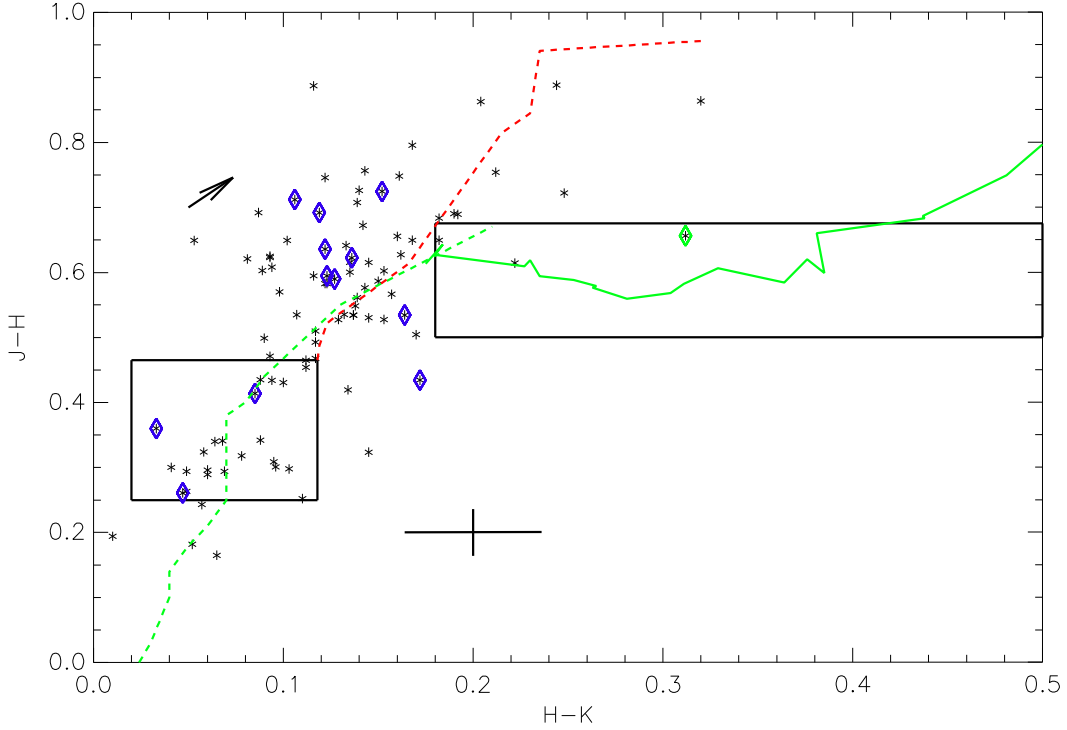}} 
\end{center}
\caption{Infrared color-color diagram of stars in the vicinity of GJ 1214 (on the plane of the sky) and with $K$-band magnitudes between 8.282 and 10.782 ($K$ = 8.782 for GJ 1214).  The horizontal and vertical black lines indicate the typical error in the colors, and the black arrow indicates the interstellar reddening vector.  The dashed green curve is the 2MASS main-sequence locus from \citet{stead2011}, the solid green curve is a main-sequence locus for later type stars, and the dashed red curve is the 2MASS giant locus.  GJ 1214 is marked with a small green diamond.  The two regions marked by black rectangles are where some reference stars were selected based on the location of the main-sequence locus.  The locations of the boxes were chosen to avoid the giant locus.  The upper right corner of the selection box on the left side of the figure is located at the colors of a G5 III star, the bluest common giant.  The twelve comparison stars that were used for relative photometry based on color and reduced proper motion criteria from \cite{lepine2011}, their proximity to the main-sequence locus, and their location in the sky (i.e. in the WFCAM FOV) are marked with small blue diamonds.  See text for further details.
\label{colors}}
\end{figure}

We downloaded $J$, $H$, and $K$ magnitudes from the 2MASS catalog \citep{skrut2006} and proper motions from the PPMXL catalog \citep{roeser2010} for all stars in the vicinity (within 30 arcmin) of GJ 1214.  We also downloaded $V$ magnitudes from the Fourth U.S. Naval Observatory CCD Astrograph Catalog \citep{zach2013}.  We imposed a magnitude cut in $K$-band so that the reference stars were no more than 0.5 mag brighter and 2 mag fainter than GJ 1214 ($K$ = 8.782).  Figure \ref{colors} is a color-color diagram of all stars that meet our magnitude criteria.  Not all these stars are actually located in the FOV, since WFCAM has four cameras, and there are gaps of 12.83 arcmin between the different cameras.    

We selected reference stars from those shown in Figure \ref{colors} based on three different criteria.  We first selected eight stars that were more likely to be M dwarf stars based on color criteria from \citet{lepine2011}.  Additional reference stars were selected based on their proximity to a 2MASS main-sequence locus from \citet{stead2011} and a locus for K7 - M9.5 spectral types \citep{cutri2003}.\footnote{http://www.pas.rochester.edu/$\sim$emamajek/memo$\_$colors.html}  Specifically, we selected stars in two regions of color space, marked by the two boxes in Figure \ref{colors}.  These regions were also selected to avoid the giant locus.\footnote{http://www.ast.leeds.ac.uk/$\sim$phy2j2s/Intrinsic$\_$Stead10.pdf}  We identified three additional stars that fit this color criteria (and that also were in the FOV).  Finally, we computed the magnitude of the proper motion for each star shown in Figure \ref{colors} (using proper motions from the PPMXL catalog).  Following \citet{lepine2011}, we computed the reduced proper motion, $H_V$, as

\be
H_V=V+5\log\mu+5.
\ee

We then applied the constraint from \citet{lepine2011} to separate M dwarfs from giants based on

\be
H_V>2.2(V-J)+2.0.
\ee

We found that all stars in Figure \ref{colors} met this criterion.  Therefore, to maximize the number of optimal comparison stars, we chose to select additional stars with the highest proper motions of the sample.  Of the five stars with the highest proper motions (excluding GJ 1214), only one was actually located within our FOV (i.e. not in a gap between cameras).  We added this star to our reference ensemble, bringing the total number of comparison stars to 12.

A preliminary visual examination of the light curve of each reference star (the flux of a given reference star divided by the total flux of the remaining reference stars) revealed that four of the twelve stars were potentially variable (i.e. the light curves displayed possible periodic fluctuations).  For completeness, we considered all 12 reference stars when generating light curves for GJ 1214.  The number of reference stars ultimately included in the light curve analysis varied between two and ten depending on which combination of references produced the lowest scatter in each set of baseline (out-of-transit) data.  

\subsection{Generating the Light Curves}
\label{lcs}
Light curves were generated from each data set by computing the relative flux ratio, i.e. the target signal divided by the total weighted signal of an ensemble of reference stars.  Due to varying weather conditions and the intrinsic variability of some of the reference stars, the reference ensemble was composed of different stars for each set of observations.  To determine which reference stars would be included in each ensemble, the rms of each reference star's signal was compared to an imposed photometric precision threshold (set to slightly different values for each night to include an adequate number of reference stars, with a typical value of 8$\times10^{-3}$).  The stars that had an rms that exceeded the threshold were excluded from the reference signal, and the remaining stars had their signals weighted (based on the rms of each individual reference signal) when computing the combined ensemble reference signal.  

Normalized light curves were generated for each photometric aperture by dividing the relative flux ratio by the median relative flux ratio measured in the out-of-transit data.  We considered the resulting out-of-transit rms scatter as measured in each aperture and for each night of observations, and we chose the aperture that gave the smallest scatter in a given night.

\begin{figure}
\begin{center}
{\includegraphics[scale=0.6, angle=90]{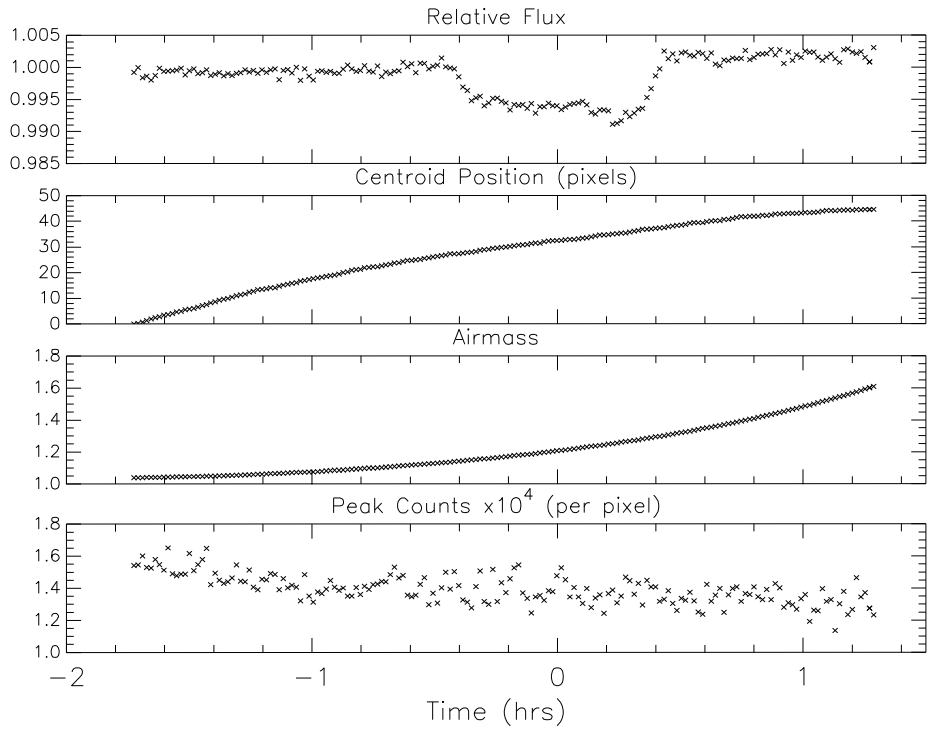}} 
\end{center}
\caption{Plots of the raw light curve (relative flux versus time) for the August 24, 2011 transit of GJ 1214b (top panel), along with parameters used for detrending the light curve (bottom panels).  The centroid position and peak counts shown are for the target.  The detrended light curve is plotted in Figure \ref{lcplot}.
\label{trends}}
\end{figure}

Each light curve was then regressed against target centroid position, airmass, and peak counts (per pixel) and a linear trend removed.  In Figure \ref{trends}, we show the light curve from August 24, 2011 prior to detrending, along with the parameters used for detrending.  This light curve has a notable negative deviation after mid-transit, which we discuss in Section \ref{systematics}.  We also tested correcting against variations in the absolute flux of the target and the variance of the flux of the target, but we found that these additional corrections resulted in a negligible change in the flux ratios.

We discarded some June 16, 2012 data due to an incorrect defocus setting and saturation (Section \ref{obs}), as well as some points that were unexplained extreme outliers ($>$ 3$\sigma$ from the mean of either the baseline data or the data at the bottom of the transit).  We discarded one data point from the August 24, 2011 observations and 28 points from the June 16, 2012 observations.  We then fit transit models to the light curve data, as described in the next section.  

\subsection{Transit Models}
\label{lcmodels}
To measure $R_p$/$R_{\star}$ we fit transit models to our seven transit light curves using the Transit Analysis Package (TAP), a publicly available IDL code \citep{gazak2012}.\footnote{http://ifa.hawaii.edu/users/zgazak/IfA/TAP.html}  TAP fits limb-darkened transit light curves using EXOFAST \citep{eastman2013} to calculate the model of \citet{mandel2002}, along with a combination of Bayesian and Markov Chain Monte Carlo (MCMC) techniques.  TAP also employs a wavelet-based likelihood function \citep{carter2009} to robustly estimate parameter uncertainties by fitting both uncorrelated (``white'') and correlated (``red'') noise.  A quadratic limb darkening law was used, so in the models described below, linear and quadratic ($\mu_{1}$ and $\mu_{2}$) limb darkening coefficients are included.  

We modeled all seven transits simultaneously, using five MCMC chains with lengths of 100,000 links each, and keeping the following parameters fixed at a single value for all transits: orbital period ($P$), inclination ($i$), scaled semi-major axis ($a$/$R_{*}$), and the limb darkening coefficients ($\mu_{1}$ and $\mu_{2}$).  (We assumed a circular orbit.)  The mid-transit time, as well as the white and red noise, was individually fitted for each transit.  To obtain $R_{p}$/$R_{*}$ we (i) fit individual transit light curves and (ii) fit the combined light curves of all transits.  The fixed parameters we used are given in Table \ref{pars}.  The $K$-band limb darkening parameters are from \citet{claret2011} for a star with $T_{\rm eff}$ = 3000 K and log($g$) = 5 and are the same as those used by \citet{narita2012}.

\begin{figure}
\begin{center}
{\includegraphics[scale=0.6, angle=90]{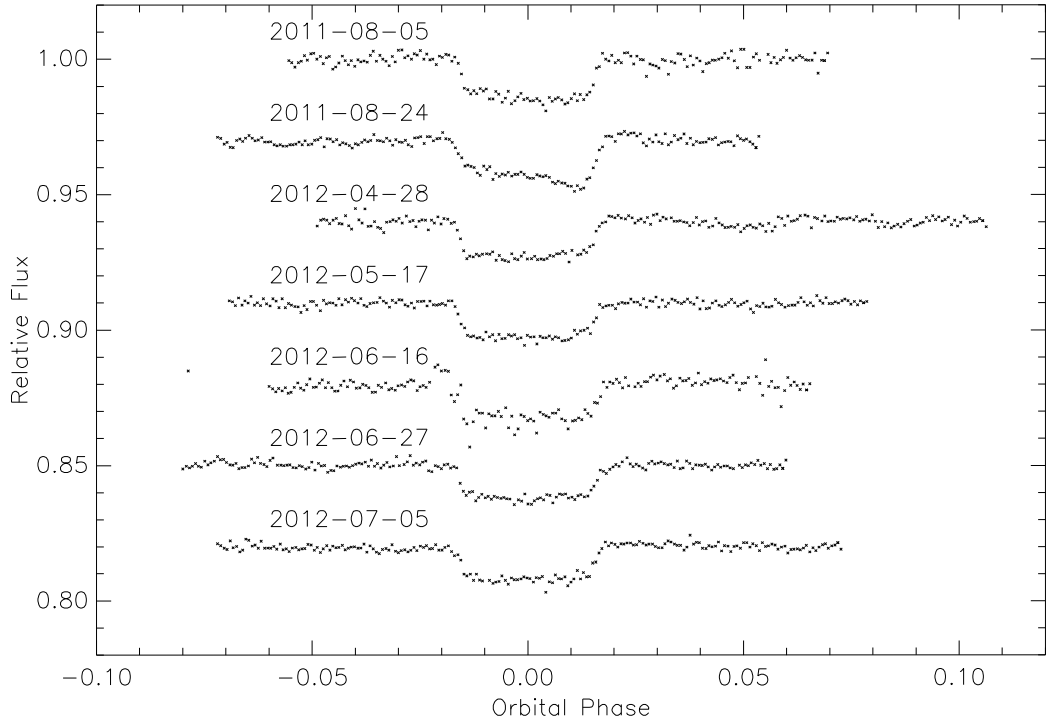}} 
\end{center}
\caption{Detrended light curves for each of the seven observed transits of GJ 1214b.  The light curves have been offset for clarity.  Each light curve has been corrected against linear trends in the target centroid position, airmass, and the peak counts in the target (per pixel).  See Figure \ref{trends} for an example of the parameters used to detrend the light curves.  
\label{lcplot}}
\end{figure}

\begin{figure}
\begin{center}
{\includegraphics[scale=0.6, angle=90]{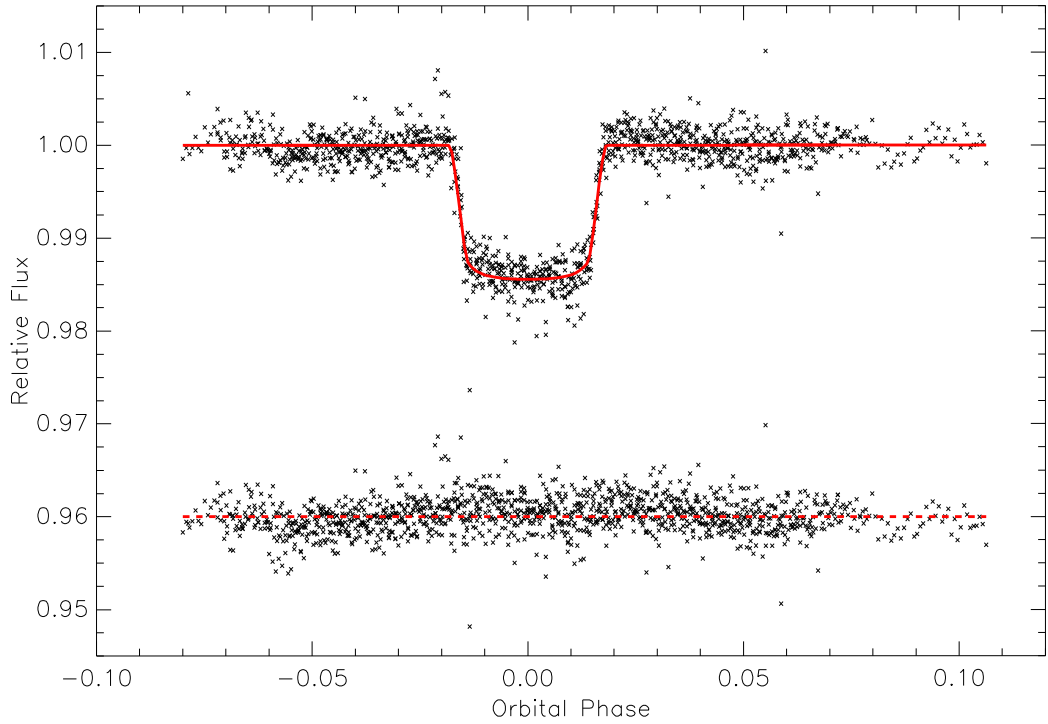}} 
\end{center}
\caption{Combined light curve (black points), best-fit model (solid red curve), and light curve residuals (offset black points) of transits of GJ 1214b.  The light curve shown is a combination of the data from all seven transits shown in Figure \ref{lcplot}.  The solid red curve is a best-fit model based on fitting a single radius ratio over all transits.  The light curve residuals after removing the model from the data are offset for clarity and have an rms of 2.0$\times10^{-3}$.  See text for further details.
\label{phaseplot}}
\end{figure}

In Figure \ref{lcplot}, we present the seven individual light curves.  Figure \ref{phaseplot} shows the combined light curve (composed by phasing each of the individual light curves) and the best-fit model for case (ii), where we fit a single $R_{p}$/$R_{*}$ to all transits.  We base our primary conclusions on the results from this case.  However, we discuss the individually measured radius ratios in Section \ref{results} and the effects of fixing versus fitting the limb darkening parameters and $a$/$R_{*}$ in further detail in Section \ref{ld}.

\section{Results}
\label{results}
In Table \ref{res} we present the best-fit radius ratio, mid-transit time, white noise, and red noise measured for each individual transit.  We also include the best-fit radius ratio based on fitting all seven transits together.  Since TAP fits for correlated (red) noise, the uncertainties on the fitted parameters should be conservative.  This was also pointed out by \citet{teske2013}, who compared results from TAP with results from other light curve fitting software.  The correlated noise is discussed in further detail in Section \ref{systematics}. 

We find that all transit times deviate from the predicted ephemeris (from \citealp{berta2012}) by less than 29 s and are consistent with a linear ephemeris.  We also find that the fitted values for the white noise are generally consistent with the rms scatter in each transit light curve (Section \ref{photo}), which suggests that the red noise may be smaller than the values measured by TAP.  We present the individual best-fit radius ratios in Figure \ref{rads}.  They are all consistent with the combined best-fit radius ratio (0.1158 $\pm$ 0.0013).  The radius ratio measured for the fifth transit observation (June 16, 2012) has the largest uncertainty, but it is still consistent with those measured from the other individual transits.  The photometry of this particular transit was of inferior quality than the other transit observations (Figure \ref{lcplot}), likely due in part to the presence of mixed thin and thick clouds throughout the observations.  

\begin{figure}
\begin{center}
{\includegraphics[scale=0.6, angle=90]{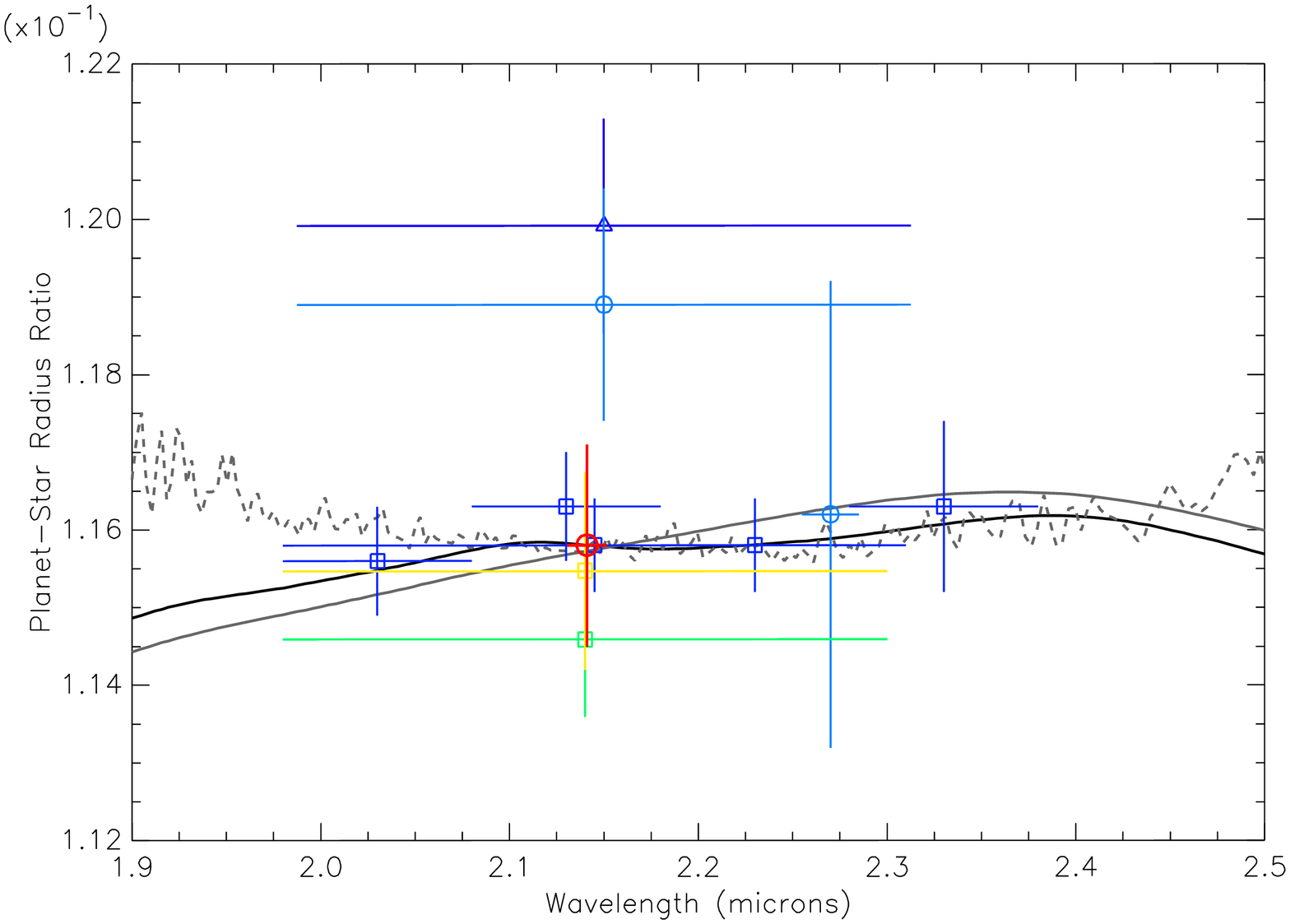}} 
\end{center}
\caption{$R_p$/$R_{\star}$ from our analysis (red circle) compared to others published in the literature.  The symbols are the same as in Figure \ref{specall}: the blue triangle is from \citet{croll2011}, blue squares are from \citet{bean2011}, blue circles are from \citet{demoo2012}, the green square is from \citet{narita2012}, and the yellow square is from \citet{narita2013}.  Vertical error bars are one standard deviation.  The horizontal error bars on each point indicate the approximate bandpass of the filter used for each observation.  The solid black and gray curves are 400 and 1000 K pure H$_2$ atmosphere models.  The models have been offset by a reference radius ratio, $R_0/R_{\star}$, derived from fitting the models to the data.  The 400 and 1000 K models were found to have the lowest and highest $\chi^2$ values [after comparing atmosphere models with different temperatures with the $K$-band data, excluding the outlying \citet{croll2011} and \citet{demoo2012} $K_s$-band data].  The 470 K 1\% H$_2$O and 99\% N$_2$ plus haze atmosphere model from \cite{howe2012} shown in Figure \ref{specall} as the dark gray curve is also shown here as a dashed gray curve.
\label{spec21}}
\end{figure}

In Figure \ref{spec21}, we present the results of our analysis along with other $K$-band measurements of the radius ratio for GJ 1214b.  We find that our best-fit radius ratio is consistent with the majority of other published values and supports a flat absorption spectrum for the atmosphere of GJ 1214b with a slope $s$ = 0.0016$\pm$0.0038  [excluding ``outlying'' points from \citet{croll2011} and \citet{demoo2012}].  We discuss this result further in Section \ref{compare}.

\section{Discussion}
\label{discuss}

\subsection{Comparison of $K$-Band Observations}
\label{compare}
\begin{figure}
\begin{center}
{\includegraphics[scale=0.6, angle=90]{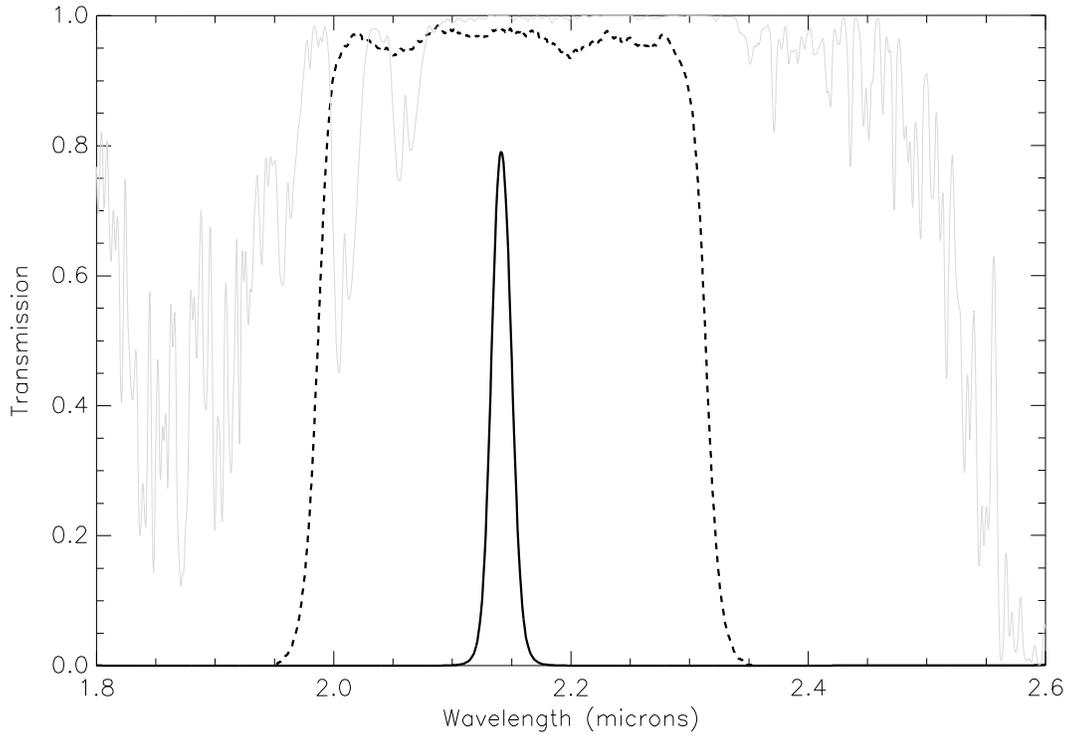}} 
\end{center}
\caption{Transmission profiles of the narrow-band H$_2$ filter (solid line), broad-band $K_s$ filter (dashed line), and atmosphere above Mauna Kea Observatory at an airmass of 1 and with a water vapor column of 1.2 mm (solid gray line).  
\label{trans}}
\end{figure}

Our measurements are inconsistent with the conclusion of \citet{croll2011}, that GJ 1214b has a low mean molecular weight atmosphere.  One $K_s$-band measurement from \citet{demoo2012} also supports the data from \citet{croll2011} and hints at a deviation from a flat spectrum (Figure \ref{spec21}).  However, the points from \citet{croll2011} and \citet{demoo2012} only differ from our derived radius ratio by 1.5$\sigma$ and 1.1$\sigma$, suggesting that their data is in fact consistent with other published $K$-band observations.  Differences between the transmission measurements could be a result of the use of different filters (i.e. our narrow-band H$_2$ filter versus their broad-band $K_s$ filter) and the correspondingly different wavelength coverage by the various groups.  Our narrow-band filter probes a small wavelength range (FWHM = 0.021 $\mu$m) and is free of telluric features (Figure \ref{trans}).  The $K_s$ transmission curve shown in Figure \ref{trans} is for the filter specifically used by \citet{croll2011} and contains some telluric absorption features.  While telluric effects should be largely removed by relative photometry, the choice of reference stars and their locations on the sky (relative to the target) can result in imperfect removal of telluric features.  The $K_s$-band also contains some stellar lines, and we discuss this further in Section \ref{photo}.  Broad-band photometry therefore can be subject to additional systematics and yield less accurate measurements than narrow-band photometry.  Ultimately, we find that $K$-band observations continue to support a high molecular weight atmosphere or a H-rich atmosphere with a cloud or haze layer over a cloud/haze-free H-rich atmosphere.

\subsection{Constraints on a Cloudy/Hazy H-Rich Atmosphere}
\label{hrich}
In the previous section, we concluded that $K$-band measurements are consistent with a flat transmission spectrum, which favors either a high molecular weight atmosphere or a H-rich atmosphere with a cloud or haze layer (Howe \& Burrows 2012).  However, the available $K$-band data does not have sufficiently high precision to discern between these two atmosphere models.  Published $B$-band data \citep[e.g.,][]{demoo2012,demoo2013,narita2013,teske2013} cannot definitively discriminate between these two scenarios either.  In this section, we assume that the atmosphere is H-dominated [following the results from Valencia et al. (2013)], and we consider what constraints can be placed on a cloudy/hazy H-rich atmosphere.  

For a H-rich atmosphere with relatively high number densities where collisions are frequent, collision-induced absorption (CIA) is a dominant source of opacity \citep[][and references therein]{borysow2002}.  Here, we consider a H-rich upper atmosphere that includes an opaque cloud or haze layer at some low altitude/high pressure.  We assume that any molecular absorbers in the atmosphere are confined beneath a cloud/haze layer, and that the atmosphere above the clouds/haze is metal-free as a result of some (photo)chemical or mixing boundary (chemopause or homopause).  If the H$_2$ envelope is sufficiently dense, it could be detected by CIA in high-precision $K$-band spectrophotometry.  This depends on temperature, so the presence or absence of CIA features is also a constraint on temperature (provided other conditions are met; see below).  The equilibrium temperature of GJ 1214b is $\sim$ 500 K \citep{charb2009}, but the upper atmosphere could be much hotter due to XUV heating from the star \citep{lammer2013}.

A condition for this model to hold is that the CIA opacity of H$_2$ dominates above the pressure altitude of any cloud/haze layer present in the atmosphere (i.e. $\tau_{H_2}$ $\gtrsim$ 1 above the chemopause or homopause).  Under this condition, we computed the minimum pressure ($P_0$) at the top of a cloud/haze layer located in an isothermal atmosphere using the following equation: 

\be
P_{0}={k_{B}T}\sqrt{\frac{1}{\sigma(T,\lambda)\sqrt{\pi R_{0} H}}}.
\ee  

$H$ is the scale height of the atmosphere, defined as

\be
H=\frac{k_BT}{\mu g},
\ee

$k_B$ is the Boltzmann constant, $T$ is the atmospheric temperature, $\mu$ is the mean molecular weight of the atmosphere, and $g$ is the planet's surface gravity.   
We assumed $T$ = 500 K, and we set the reference planetary radius $R_{0}$ = 2.68 $R_{\oplus}$.  We used H$_2$-H$_2$ CIA opacities ($\sigma$) from \citet{borysow2002} and computed the scale height $H$ using a mean molecular weight $\mu$ = 2 and a surface gravity $g$ = 8.95 m s$^{-2}$.  From this, we computed the minimum pressure $P_0$ at the top of the cloud or aerosol layer to be 60 mbar (at $\lambda$ $\sim$ 2.14 $\mu$m).  This minimum pressure falls within the pressure range of 0.001$-$100 mbar considered by \citet{howe2012}. 

We then calculated the effective $R_p$/$R_{\star}$ versus wavelength, following a similar procedure as in \citet{howe2012}:

\be
\frac{R_{p}}{R_{\star}}=\frac{R_{0}}{R_{\star}}+\frac{H}{2R_{\star}}\ln\left[\frac{\sigma(T,\lambda)}{\sigma_{0}}\right].
\ee

Here, $R_0$ is the radius at the wavelength where $\sigma$ = $\sigma_0$ = 2.78$\times10^{-6}$ cm$^{-1}$ amagat$^{-2}$ at  $\lambda$ $\sim$ 2.14 $\mu$m [from the 500 K table from \citet{borysow2002}].   Because H$_2$ absorption is collisionally-induced, it depends on the square of the number density ($n^2$, where $n$ = number density of absorbers) and this leads to an additional factor of 1/2 in Eqn. 6 [compared to the equation defined in \citet{howe2012}].  

Under the same conditions assumed above and using $R_{\star}$ = 0.211 $R_{\odot}$ \citep{charb2009}, we calculated the radius ratio for temperatures of 400, 500, 600, 700, 800, 900, and 1000 K.  We compared the observed radius ratios to the predicted radius ratios by calculating $\chi^2$ and identified the best-fit model (based on the minimum $\chi^2$).  Although the $K_s$-band data from \citet{croll2011} and \citet{demoo2012} are marginally consistent with other $K$-band data within measurement uncertainties, we still considered them to be unexplained outliers and excluded them from the analysis (Section \ref{compare}).  

We find the 400 K model had the smallest $\chi^2$, while the 1000 K model yielded the largest $\chi^2$.  We show the 400 and 1000 K models in Figure \ref{spec21}.  As the temperature increases, the scale height increases, leading to additional absorption and a larger apparent planet radius; however, the features also become washed out at higher temperatures.  We computed $\Delta\chi^2$ (relative to the minimum $\chi^2$ at 400 K) and found that the deviation between the data and models increases with increasing temperature.  However, $\Delta\chi^2$ between the 400 and 1000 K models is only $\sim$ 0.72, since $\chi^2$ is just 2$-$3 for all models.  We conclude that from the available data, we cannot exclude higher temperature atmospheres ($T$ $>$ 400 K) with any confidence ($p<$ 0.01). 

Considering the capabilities of future missions like the \emph{James Webb Space Telescope} ($JWST$) for high precision infrared spectroscopy, the atmosphere models were compared to both the real data (using the actual measurement uncertainties) as well as an artificial data set, consisting of the actual measurements with reduced errors.  We defined artificial errors so that the median error over the $K$-band data = 1$\times10^{-4}$ [based on the precision achieved by \citet{fraine2013} from $Spitzer$ measurements of GJ 1214b at 4.5 $\mu$m].  For comparison, the median value of the actual errors is 9.9$\times10^{-4}$.  Based on the artificial high-precision data, we again find that the 400 K model has the smallest $\chi^2$, and that $\Delta\chi^2$ between the 400 and 1000 K models is 70.2, which would allow us to exclude atmospheres with $T$ $\ge$ 800 K with $>$ 99.7\% confidence (3$\sigma$) assuming that we had such high precision data.  We find that a temperature of $\le$ 400 K is preferred for a pure H$_2$ upper atmosphere, but significantly higher precision data (as well as more data in general) is needed to confidently exclude higher temperature atmospheres.  

Ultimately, while we find no evidence for deviation from a flat spectrum, a thin upper atmosphere ($\le$ 60 mbar) dominated by H$_2$ cannot be excluded.  The possibility also remains that there is simply no (or very little) H in the (upper) atmosphere.

\subsection{Variability in the Stellar Spectrum}
\label{photo}

Fluorescent H$_2$ emission lines have been observed in the spectra of four planet-hosting M dwarf stars, but not in GJ 1214 \citep{france2013}.  These lines are produced by photoexcitation by Ly$\alpha$ photons, and their detection indicates the presence of a 2000 $-$ 4000 K molecular gas \citep[][and references therein]{france2013}.  The H$_2$ filter is designed specifically to observe such lines, making it possible for us to also probe the stellar atmosphere with our observations.  While \citet{france2013} do not detect Ly$\alpha$ emission in GJ 1214, they speculate that this is because the neutral H in GJ 1214's atmosphere is instead contained in H$_2$ rather than H.  Notably, GJ 1214 is the coolest star ($T_{\rm eff}$ $\sim$ 3250 K; Anglada-Escud{\'e} et al. 2013) in their sample, and H recombines at cool temperatures, so it is plausible that molecular H is present in GJ 1214's atmosphere.  This motivated us to look for H$_2$ emission in GJ 1214.
  
\begin{figure}
\begin{center}
{\includegraphics[scale=0.6, angle=90]{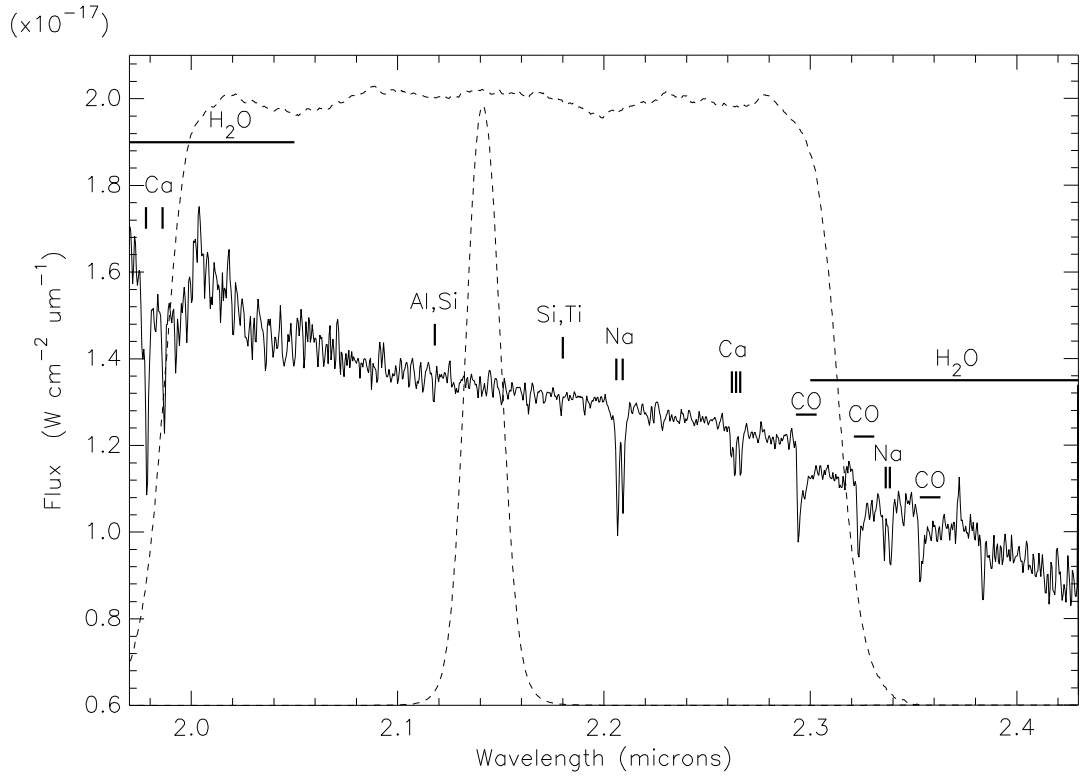}} 
\end{center}
\caption{$K$-band spectrum of GJ 1214 from \citet{rojas2012}.  Pronounced stellar absorption lines are labeled.  The profiles of the narrow-band H$_2$ filter and the $K_s$ filter are illustrated by the dashed curves. 
\label{gj}}
\end{figure}

We examined a $K$-band spectrum of GJ 1214 from \citet{rojas2012}, shown in Figure \ref{gj}.  We find no evidence of an H$_2$ emission feature around $\sim$ 2.14 $\mu$m in GJ 1214's spectrum.  The H$_2$ line is close to the Brackett $\gamma$ line (2.16 $\mu$m), which is produced in T Tauri stars by recombining magnetospheric gas \citep[e.g.,][]{ham1988}.  However, we see no evidence of the Brackett $\gamma$ line in the spectrum of GJ 1214.  Indeed, the H$_2$ bandpass is free of any obvious stellar lines.  The $K_s$ filter does include several stellar absorption lines, but assuming that these features are only present in the star (and are not telluric) and that they do not vary, these will not affect the  photometry.

\begin{figure}
\begin{center}
{\includegraphics[scale=0.6, angle=90]{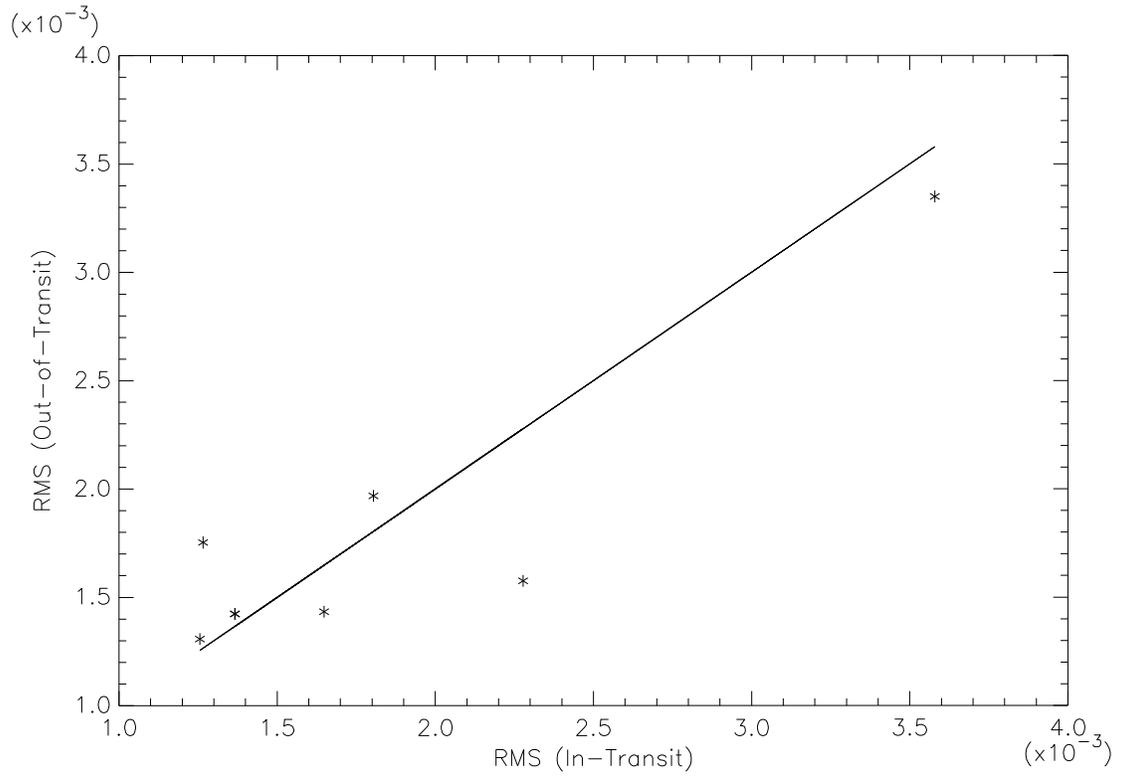}} 
\end{center}
\caption{The in-transit rms versus out-of-transit rms for each observed transit.  The solid line illustrated equality between the in- and out-of-transit rms values.
\label{rms}}
\end{figure}

Due to the lower temperatures H$_2$ may predominate in star spots and may play an important role in the formation and evolution of spots \citep[e.g.,][]{jaeggli2012}.  Therefore, we also looked for evidence of H$_2$ variability, or patchiness in the stellar disk in the H$_2$ bandpass (either bright emission or dark absorption) which produces variability when transited by the planet (i.e. ``H$_2$-spots'').\footnote{While we find no strong evidence of spot-crossing events in any of our light curves, we discuss a possible $bright$ spot in Section \ref{systematics}.}  We compared observations taken in-transit (when the time-varying part of the star is blocked by the planet's disk) with those taken out-of-transit (when only the star light is visible).  Specifically, we compared the rms scatter (of the light curve residuals, computed by subtracting the best-fit model from the data; Section \ref{lcmodels}) between the in-transit and out-of-transit windows for each of the seven individual transits.  The resulting rms values are presented in Figure \ref{rms}.  We computed $r^2$ = 0.80 between the in- and out-of-transit rms, which indicates a strong linear correlation between the two parameters.  That a strong linear correlation exists suggests that the in- and out-of-transit measurements are dominated by the same source of variation, i.e. photometric errors as opposed to a source of variation associated with the transit alone.  From this, we conclude that if H$_2$ is present and absorbing or emitting in the stellar and/or planetary atmosphere, it is either not variable or its variations are not detectable in our data.  It is likely that the vibrational transitions such as the 1-0 S(1) transition are too weak to be seen in a standard $K$-band spectrum \citep[but instead might be seen through a linearly polarized spectrum;][]{white2011}.  
 
The H$_2$ 1-0 S(1) line is also used to study molecular outflows from protostars \citep[e.g.,][]{garcia2013}.  We considered the possibility of detecting H$_2$ ``outflows'' from the atmosphere of the planet (driven by UV heating), but we see no persistent deviations from the standard transit light curve models that require explanation by a shock or wind (i.e. from some massive atmospheric escape from the planet).

\subsection{Effects of Systematics}
\label{systematics}
Systematics, whether astrophysical or instrumental in nature, can significantly affect the derivation of light curve parameters.  In this section, we describe potential sources of systematics in our light curves and the effect they have on our measured radius ratios.

\subsubsection{Astrophysical Sources}

\begin{figure}
\begin{center}
{\includegraphics[scale=0.6, angle=90]{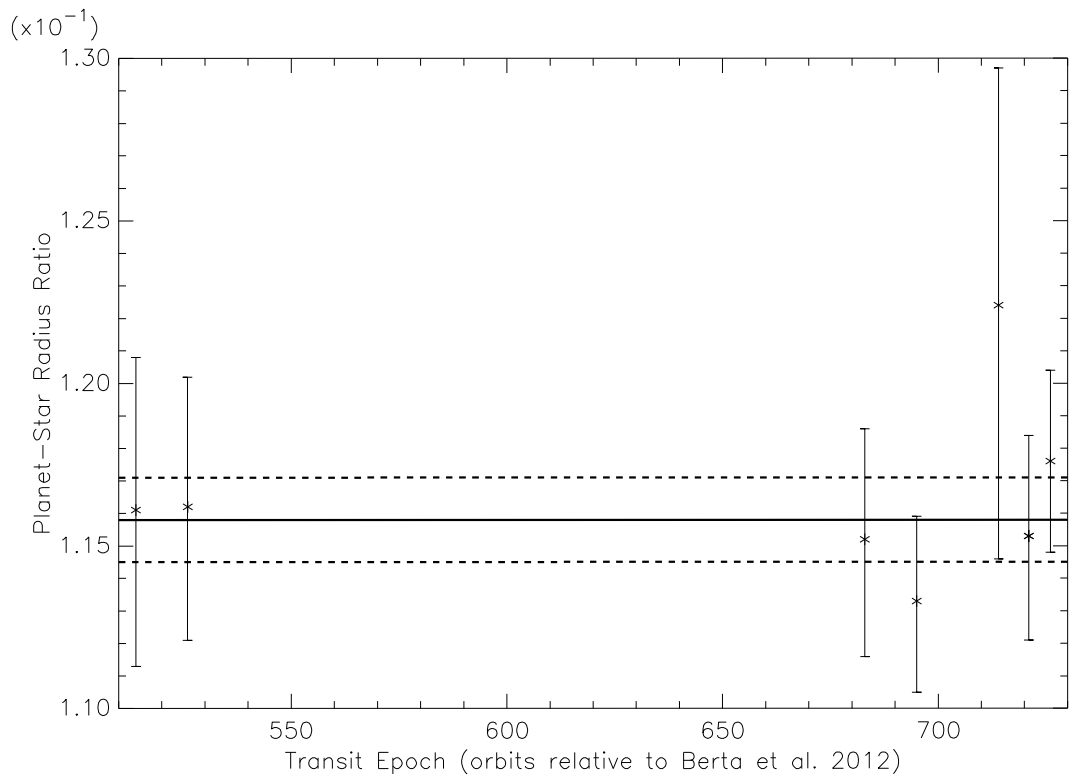}} 
\end{center}
\caption{Best-fit radius ratios derived from TAP models fit to the data.  We show the best-fit radius ratios and their corresponding 1$\sigma$ uncertainties from fitting the seven transits separately versus transit epoch based on the ephemeris from \citet{berta2012}.  The solid line is the best-fit radius ratio from fitting all seven transits together (with the $\pm$1$\sigma$ uncertainties shown as dashed lines).
\label{rads}}
\end{figure}

Intrinsic variability of the host star can affect measurements of the radius ratio, and GJ 1214 has been shown to be variable at red ($\sim$0.8 $\mu$m) wavelengths at the 1-2\% level over $\sim$ 1-2 year timescales \citep[][]{carter2011}.  While we see no obvious visual evidence of stellar flares or star spots\footnote{When transited by the planet, the presence of a star spot would result in a brightening event in the light curve.  When spots are present but are not transited by the planet, they reduce the overall brightness of the star and thus produce a deeper transit.} in our light curves (Figure \ref{lcplot}), in principle it is possible to use the radius ratios measured for the individual transits to estimate how much the spot coverage changed with time.  Theoretically, the shallowest observed transit should correspond to the stellar surface being nearly free of spots, or at least having minimal spot coverage.  As presented in Table \ref{res} and shown in Figure \ref{rads}, the measured radius ratios do not vary significantly with transit epoch.  Between the photometric quality of the data and the magnitude of the errors derived for the radius ratios, we cannot robustly identify a minimum epoch of stellar activity based on the measured radius ratios.  Based on the rms values presented in Figure \ref{rms}, our data suggest variability at a level of $\sim$ 1.5$\times10^{-3}$ (albeit only on hour-long timescales).  We also find that GJ 1214 is variable in the $K$-band at a level of $\sim$ 0.5\% over longer timescales (from August 2011 to July 2012 or $\sim$ 1 year), based on the target-to-reference flux ratio for the most photometrically stable reference star in our sample.  This is consistent with the results from \citet{carter2011}, since stellar variability in our passband should be less than the 1-2\% measured by \citet{carter2011} due to observing in a redder passband.  Although such variability does affect the measurement of radius ratios, since our radius ratios did not vary significantly (within errors), we consider stellar variability to have a negligible effect on our conclusions. 

\begin{figure}
\begin{center}
{\includegraphics[scale=0.6, angle=90]{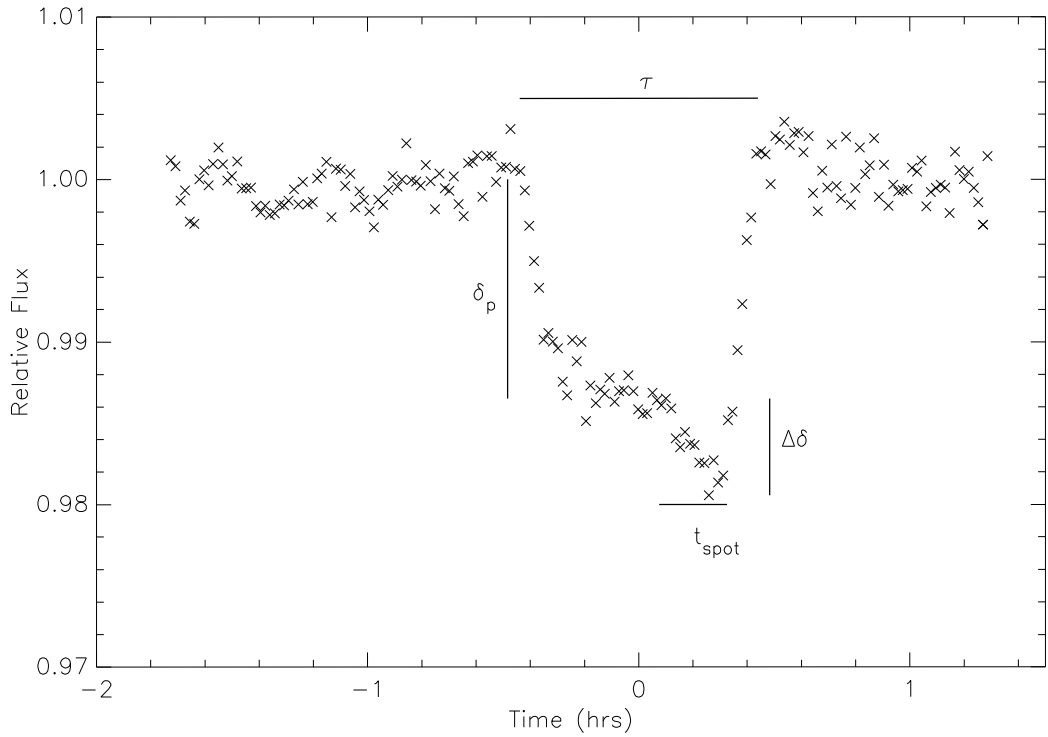}} 
\end{center}
\caption{Light curve for the August 24, 2011 transit of GJ 1214b.  The intervals for the transit duration $\tau$ and the crossing time of a hypothetical bright spot $t_{\rm spot}$ are marked with horizontal black lines.  The vertical black lines indicate the depth of the planetary transit ($\delta_p$) and the change in the transit depth due to the anomalous feature ($\Delta\delta$).
\label{spot}}
\end{figure}

There is the possibility that spot-crossing events did occur in one or more of our observed transits, but we simply cannot identify them by eye because they cause brightenings at or below the level of our photometric precision.  If we consider the red noise (Table \ref{res}) as a measure of the effects of spot-crossing events, we find that the August 24, 2011 and June 16, 2012 transits have the highest levels of red noise.  The June 16, 2012 transit had poorer weather conditions, which is likely the cause of the higher red noise in that light curve, as no evidence of a spot-crossing event is seen in that light curve.  However, the August 24, 2011 transit, which had photometric conditions, does have an anomalous feature in its light curve (Figure \ref{spot}) where the transit depth appears to increase significantly immediately before egress.  The anomalous feature in the August 24, 2011 transit does not appear to be a (dark) spot crossing event, since the transit depth before the feature occurred is consistent (i.e. within 1$\sigma$) with the transit depth measured for the combined light curve (over all seven transits).  This suggests that the true transit depth should be based on the first part of the transit, rather than the first part of the transit being a spot-crossing event. 

To determine if this anomalous feature was produced by reference star variability, we constructed alternative light curves using different reference stars.  We found that the feature was present in all cases.  We also searched the literature to see if any other observations of this specific transit had been published, but we only found an observation of the transit following this one by \citet{harpsoe2013}.  Their light curves do not appear to contain any particularly anomalous features, which supports the idea of minimal (or at least un-transited) spots at the time of the observations.  

We also considered that the feature may instead be due to the planet passing over a $bright$ spot on the star, which would result in an increase in the transit depth, consistent with our observations.  However, the presence of a bright spot will also cause a $decrease$ in the unocculted transit depth.  As illustrated in Figure \ref{rads}, the transit depths measured for all transits are consistent within 1$\sigma$.  This suggests that if the anomalous feature is a result of a spot-crossing event, the spot is sufficiently compact/faint that it does not produce an unocculted spot effect that we would detect.  We do not find evidence of spot-crossing events in our other transit observations, but we explore the bright spot hypothesis in more detail here.    

Assuming a circular spot with radius $R_{\rm spot}$, we estimated the size of such a spot based on the duration of the spot crossing,

\be
\frac{R_{\rm spot}}{R_*}=\frac{t_{\rm spot}}{\tau}\sqrt{1-b^2}.
\ee

The transit duration ($\tau$) is 52.73 min \citep{carter2011}, and the spot-crossing time ($t_{\rm spot}$) was estimated from the light curve to be 15 min (Figure \ref{spot}).  Using an impact parameter ($b$) of 0.28 \citep{bean2011}, we computed a spot-star radius ratio of 0.27 (notably over two times larger than $R_p$/$R_{\star}$).  We then computed how much the transit depth would increase due to a bright spot (relative to the no spot case) from

\be
\delta_f=B\left(\frac{R_{\rm spot}}{R_*}\right)^2.
\ee

This equation is applicable because our hypothetical spot is much larger than the planet.  The increase in the transit depth during the anomalous event relative to the depth prior to the anomalous event is $\Delta\delta$ = 0.0059 (Figure \ref{spot}).  The transit depth ($\delta_p$) is 0.0135.  $B$ is the brightness enhancement of the spot relative to the rest of the stellar disk, which we computed based on the relative depth of the anomalous feature, or $B$ = $\Delta\delta$/$\delta_p$ = 0.44.  From this, we calculate $\delta_f$ = 0.033.  This translates to a change in the transit depth due to the unocculted bright spot of $\delta_f$$\times$$\delta_p$ = 4.4$\times10^{-4}$, which is much smaller than our typical photometric precision of 1.7$\times10^{-3}$ (for a single transit).  It is also consistent with GJ 1214 being variable on the order of 1-2\% \citep{carter2011}.  We conclude that a single transient bright spot about twice the size of the planet's disk could explain the transit of August 24, 2011.  Regardless of the source of the feature, TAP recognized it as a systematic and derived a radius ratio that is consistent with the radius ratios measured from the other transits (Figure \ref{rads}).  

\subsubsection{Instrumental Sources}

Besides astrophysical systematics, we also considered the effect of instrumental systematics such as a nonlinear detector on our measured radius ratios.  The WFCAM detector is linear to $<$ 1\% up to about 40,000 counts per pixel.\footnote{http://apm49.ast.cam.ac.uk/surveys-projects/wfcam/technical/linearity}  To ensure that we avoided the defined non-linear regime, we measured the peak counts in the target and each reference star.  We found that all stars remained below $\sim$22,000 counts (per pixel) in all observations.  However, the possibility remains that lower non-linearity is present at lower counts (Section \ref{obs}).  To check this, we derived a ``variance coefficient'' ($k$) for each reference star and for each night based on the following equation,

\be
F_{\rm observed}=F_{\rm actual}-kF_{\rm observed}^2.
\ee

Here, $F_{\rm observed}$ is the normalized flux ratio we measured for each reference star, $F_{\rm actual}$ is what the ideal flux ratio should be in the absence of systematics (i.e. $F_{\rm actual}$ $\equiv$ 1), $k$ is defined as the variance coefficient, and $F_{\rm observed}^2$ is the variance of the observed flux ratio.  After deriving $k$ for each reference star and for each night, we found that on most nights, $k$ $<<$ 1.1$\times10^{-3}$.  Thus, we conclude that the effect of non-linearity is smaller than our photometric precision (typically 1.7$\times10^{-3}$ for a single transit).  Only on two nights did the flux ratios have significant deviations:  during the fifth transit (June 16, 2012) and sixth transit (June 27, 2012).  That the fifth transit shows signs of non-linearity is consistent with the observations, since the stellar image was insufficiently defocused to avoid high counts.  We have no explanation as to why the sixth transit is potentially affected by non-linearity.  Given that the measured transit depth is consistent with those measured from the other individual transits, we conclude that any low-level non-linearity that might be present and a source of systematics has a minimal effect on our photometry.  

\subsection{Treatment of Transit Model Parameters}
\label{ld}
Finally, we considered how limitations in our knowledge of stellar properties affect the derived light curve parameters.  In particular, there is a degeneracy between limb darkening and $a$/$R_{*}$, which in turn affects estimates of the stellar density, orbital eccentricity, and impact parameter.  Furthermore, as \citet{berta2012} point out, inaccurate treatment of limb darkening could introduce false absorption features into the transmission spectrum.  Therefore, the accurate treatment of limb darkening is critical for precisely measuring light curve parameters.  

In the analyses described above, we held limb darkening coefficients and $a$/$R_*$ fixed.  This decision was based on the fact that GJ 1214b's transit duration is only 52 min and the ingress/egress events last 6 min, compared to our cadence of $\sim$ 1 min.  Having a relatively small number of data points during ingress and egress makes it difficult to accurately fit the limb darkening.  However, \citet{csizmadia2013} argue that stellar limb darkening parameters should be fitted and not fixed in order to derive high-precision light curve parameters.  Thus, we used TAP to fit additional models to our data, with (1) limb darkening as a free parameter and $a$/$R_*$ held fixed, (2) limb darkening held fixed and $a$/$R_*$ as a free parameter, and (3) both limb darkening and $a$/$R_*$ as free parameters.  In all cases, we fit a single radius ratio over all seven transits.  For the models where limb darkening was a free parameter (1 and 3), we derive linear and quadratic limb darkening coefficients of (0.089$\pm$0.018, -0.131$\pm$0.018) and (0.090$\pm$0.018, -0.130$\pm$0.018).  Compared to the fixed values that we used, (0.0475, 0.3502), we find that the fitted linear coefficient is consistent with the fixed coefficient within 3$\sigma$, but the fitted quadratic coefficient differs from the fixed coefficient by more than 26$\sigma$.  The derived value for $a$/$R_*$ from model 2 (15.111$^{+0.081}_{-0.080}$) is consistent with the fixed value we used (14.975) within 1.7$\sigma$.  For model 3, we found $a$/$R_*$ = 15.410 $\pm$ 0.080, which differs from the fixed value by 5.5$\sigma$.  Despite differences between some of the fixed and fitted parameters, we find that for all models, the derived radius ratios are consistent within 1$\sigma$.  Specifically, the measured radius ratios for the three cases described above are: (1) $R_p/R_{\star}$ = 0.1174$\pm$0.0015 when limb darkening is free and $a$/$R_*$ is fixed, (2) $R_p/R_{\star}$ = 0.1157$\pm$0.0013 when limb darkening is fixed and $a$/$R_*$ is free, and (3) $R_p/R_{\star}$ = 0.1175$\pm$0.0013 when both limb darkening and $a$/$R_*$ are free.  Recall that our base model yielded $R_p/R_{\star}$ = 0.1158$\pm$0.0013 when keeping both limb darkening and $a$/$R_*$ fixed.  Since all radius ratios are consistent within 1$\sigma$, this indicates that fixing limb darkening as well as $a$/$R_*$ did not significantly affect our results or conclusions.

\section{Conclusions}
\label{conc}
In this paper, we presented results from seven transit observations of GJ 1214b acquired in a narrow-band H$_2$ filter.  Our analysis included a thorough technique for selecting reference stars and incorporated light curve fits with red (correlated) noise.  We measured a radius ratio of 0.1158$\pm$0.0013 when fitting the data from all seven transits together.  This radius ratio is consistent with previous $K$-band measurements, including those from \citet{croll2011} and \citet{demoo2012} which differ from ours by only 1.5$\sigma$ and 1.1$\sigma$.  We conclude that all $K$-band data support a flat absorption spectrum for GJ 1214b, which suggests that either a high mean molecular weight atmosphere or a H-rich atmosphere with a cloud or haze layer is favored \citep{howe2012}.

Since \citet{valencia2013} find that the bulk amount of H/He in GJ 1214b's volatile envelope is nonzero and could be as much as 7\% by mass, we explored the scenario where GJ 1214b has a H$_2$-rich envelope and heavy elements are sequestered below a cloud or aerosol layer.  After comparing models of a pure H$_2$ atmosphere with $K$-band observations, we find that we cannot exclude the possibility of a H$_2$-rich upper atmosphere.  It is difficult to disentangle different plausible atmosphere models in the $K$-band given the precision of available data and that the temperature of the upper atmosphere is otherwise unconstrained.  We suggest that additional high-precision spectroscopic observations (from space) across the $K$-band would be most useful.  High-precision measurements at short optical wavelengths ($<$ 0.7 $\mu$m) would also be helpful in searching for evidence of Rayleigh scattering \citep[e.g.][]{demoo2013,narita2013}, which would support a (cloudy/hazy) H-rich atmosphere for GJ 1214b \citep{howe2012}.

Finally, we explored variability due to H$_2$ in the stellar spectrum and investigated the effects of different systematics on our results, such as star spots, CCD non-linearity, and limitations in our knowledge of light curve parameters.  Overall, we conclude that systematics did not significantly affect our photometry and that our results are robust to these effects.

While there is ample spectroscopic and photometric data available for this planet, much of the data comes from different instruments as well as analyses in addition to suffering from poor precision.  This in itself introduces an additional systematic when comparing the data to atmosphere models.  While beyond the scope of this paper, we conclude that a uniform analysis of all public data for GJ 1214b would be useful for establishing more robust limits on atmosphere models.  

\acknowledgements  We are grateful to Mike Irwin and Mike Read for assisting us in accessing the WFCAM data and illumination tables.  We thank Norio Narita for helpful discussions on GJ 1214b.  We also thank Paul Wilson for providing data prior to publication.  We acknowledge the anonymous referee for helping us improve this paper.  This research was supported by NASA grants NNX10AI90G (Astrobiology: Exobiology \& Evolutionary Biology) and NNX11AC33G (Origins of Solar Systems) to EG.  The United Kingdom Infrared Telescope is operated by the Joint Astronomy Centre on behalf of the Science and Technology Facilities Council of the U.K.  This research has made use of the VizieR catalogue access tool, CDS, Strasbourg, France. 

\clearpage

\begin{deluxetable}{cccccc} 
\tabletypesize{\small}
\tablewidth{0pt}
\tablecaption{Transit Observations \label{obstable}}
\tablehead{
\colhead{Date (UT)} & \colhead{Start Time (UT)} & \colhead{End Time (UT)} & \colhead{Airmass Start} & \colhead{Airmass End} & \colhead{Notes}}
\startdata
2011-08-05 & 07:03:04.3 & 10:03:04.3 & 1.04 & 1.58 & (a) \\
2011-08-24 & 05:50:47.0 & 08:51:30.2 & 1.04 & 1.61 & \\
2012-04-28 & 09:18:43.2 & 13:01:55.2 & 1.89 & 1.03 & \\
2012-05-17 & 07:57:13.0 & 11:30:02.9 & 1.98 & 1.04 & \\
2012-06-16 & 08:21:33.1 & 11:50:12.5 & 1.13 & 1.17 & (b) \\
2012-06-27 & 09:51:50.4 & 13:13:09.1 & 1.05 & 2.01 & \\
2012-07-05 & 07:43:14.9 & 11:11:19.7 & 1.07 & 1.29 & (c) \\
\enddata
\tablenotetext{(a)}{Conditions were windy throughout the observations.}
\tablenotetext{(b)}{Conditions were mixed (thin/thick clouds), and an incorrect focus setting was used in the beginning of the observations (resulting in saturation), so the first part of the observations was discarded.  See Section \ref{analysis} for further details.}
\tablenotetext{(c)}{Conditions were windy through the first half of the observations.}
\end{deluxetable}
\clearpage

\begin{deluxetable}{ccc} 
\tabletypesize{\small}
\tablewidth{0pt}
\tablecaption{Fixed Transit Model Parameters$^{a}$ \label{pars}}
\tablehead{
\colhead{Parameter} & \colhead{Value} & \colhead{Reference}}
\startdata
$P$ (days) & 1.58040481 & \citet{bean2011} \\
$i$ (deg) & 88.94 & \citet{bean2011} \\
$a$/$R_{*}$ & 14.9749 & \citet{bean2011} \\
$\mu_{1}$$^b$ & 0.0475 & \citet{claret2011} \\
$\mu_{2}$$^b$ & 0.3502 & \citet{claret2011} \\
\enddata
\tablenotetext{(a)}{A circular orbit is assumed.}
\tablenotetext{(b)}{The linear and quadratic limb darkening coefficients are for the $K$-band.}
\end{deluxetable}
\clearpage

\begin{deluxetable}{ccccc}
\tabletypesize{\small}
\tablewidth{0pt}
\tablecaption{Best-Fit Model Parameters \label{res}}
\tablehead{
\colhead{Date (UT)} & \colhead{T$_{\rm mid}$ (JD-2450000)} & \colhead{White Noise} & \colhead{Red Noise} & \colhead{$R_{p}/R_{*}$$^{a}$}}
\startdata
2011-08-05 & 5778.84962 $^{+0.00025}_{-0.00026}$ & 0.00122 $^{+0.00015}_{-0.00017}$ & 0.0093 $^{+0.0026}_{-0.0023}$ & 0.1161 $^{+0.0047}_{-0.0048}$ \\
2011-08-24 & 5797.81581 $^{+0.00024}_{-0.00024}$ & 0.00048 $^{+0.00017}_{-0.00026}$ & 0.0109 $^{+0.0015}_{-0.0015}$ & 0.1162 $^{+0.0040}_{-0.0041}$ \\
2012-04-28 & 6045.93714 $^{+0.00024}_{-0.00024}$ & 0.00106 $^{+0.00012}_{-0.00013}$ & 0.0084 $^{+0.0020}_{-0.0017}$ & 0.1152 $^{+0.0034}_{-0.0036}$ \\
2012-05-17 & 6064.90099 $^{+0.00018}_{-0.00018}$ & 0.00104 $^{+0.00010}_{-0.00011}$ & 0.0050 $^{+0.0020}_{-0.0019}$ & 0.1133 $^{+0.0026}_{-0.0028}$ \\
2012-06-16 & 6094.92825 $^{+0.00046}_{-0.00044}$ & 0.00169 $^{+0.00038}_{-0.00046}$ & 0.0174 $^{+0.0053}_{-0.0057}$ & 0.1224 $^{+0.0073}_{-0.0078}$ \\
2012-06-27 & 6105.99128 $^{+0.00022}_{-0.00021}$ & 0.00090 $^{+0.00013}_{-0.00015}$ & 0.0076 $^{+0.0021}_{-0.0020}$ & 0.1153 $^{+0.0031}_{-0.0032}$ \\
2012-07-05 & 6113.89391 $^{+0.00019}_{-0.00019}$ & 0.00109 $^{+0.00010}_{-0.00010}$ & 0.0059 $^{+0.0019}_{-0.0016}$ & 0.1176 $^{+0.0028}_{-0.0028}$ \\
\enddata
\tablenotetext{(a)}{The best-fit radius ratio from fitting all seven transits together is 0.1158 $\pm$0.0013.  The best-fit radius ratios from fitting the seven transits separately are shown in the fifth column and are also shown in Figure \ref{rads}.  See text for further details.}
\end{deluxetable}
\clearpage

\clearpage

\clearpage

\clearpage

\clearpage

\clearpage

\clearpage

\clearpage

\clearpage


\clearpage

\clearpage

\clearpage

\end{document}